\RequirePackage{fix-cm}

\documentclass[smallextended]{svjour3}       
\smartqed  
\usepackage{graphicx}
\usepackage{url} 
\usepackage{lineno}
\usepackage{amsfonts}
\usepackage{amsmath}
\usepackage{textcomp}
\usepackage{bm}
\usepackage{soul}
\usepackage[numbers,sort&compress]{natbib}
\usepackage{color}
\begin{document}

\graphicspath{
               {Figures/}
              }

\title{Markov Chain Monte Carlo with Neural Network Surrogates: Application to Contaminant Source Identification
}

\titlerunning{CNN Surrogates for Source Identification}        

\author{Zitong Zhou         \and
        Daniel M. Tartakovsky$^*$ 
}


\institute{Z. Zhou \at
             Department of Energy Resources Engineering, Stanford University, Stanford, CA, USA
             \and
           D. M. Tartakovsky \at
              Department of Energy Resources Engineering, Stanford University, Stanford, CA, USA \\
              \email{tartakovsky@stanford.edu}           
}


\maketitle

\begin{abstract}
Subsurface remediation often involves reconstruction of contaminant release history from sparse observations of solute concentration.  Markov Chain Monte Carlo (MCMC), the most accurate and general method for this task, is rarely used in practice because of its high computational cost associated with multiple solves of contaminant transport equations. We propose an adaptive MCMC method, in which a transport model is replaced with a fast and accurate surrogate model in the form of a deep convolutional neural network (CNN). The CNN-based surrogate is trained on a small number of the transport model runs based on the prior knowledge of the unknown release history.  Thus reduced computational cost allows one to diminish the sampling error associated with construction of the  approximate likelihood function. As all MCMC strategies for source identification, our method has an added advantage of quantifying predictive uncertainty and accounting for measurement errors.  Our numerical experiments demonstrate the accuracy comparable to that of MCMC with the forward transport model, which is obtained at a fraction of the computational cost of the latter.
\end{abstract}

\section{Introduction}
\label{intro}

Identification of contaminant release history in groundwater plays an important role in regulatory efforts and design of remedial actions. Such efforts rely on measurements of solute concentrations collected at a few locations (pumping or observation wells) in an aquifer. Data collection can take place at discrete times and is often plagued by measurement errors. A release history is estimated by matching these data to predictions of a solute transport model, an inverse modeling procedure that is typically ill-posed. 
Alternative strategies for solving this inverse problem~\cite{amirabdollahian2013identification, zhou2014inverse, rajabi2018model, barajas2019efficient} fall into two categories: deterministic and probabilistic. Deterministic methods include least squares regression~\cite{white2015nonlinear} and hybrid optimization with a genetic algorithm \cite{ayvaz2016hybrid, leichombam2018new}. They provide a ``best'' estimate of the contaminant release history, without quantifying the uncertainty inevitable in such predictions.

Probabilistic methods, e.g., data assimilation via extended and ensemble Kalman filters \cite{xu2016joint, xu2018simultaneous} and Bayesian inference based on Markov chain Monte Carlo or MCMC \cite{gamerman2006markov}, overcome this shortcoming. Kalman filters are relatively fast but do not generalize to strongly nonlinear problems, sometimes exhibiting inconsistency between updated parameters and observed states \cite{chaudhuri2018iterative}. Particle filters and MCMC are exact even for nonlinear systems but are computationally expensive, and often prohibitively so. Increased efficiency of MCMC with a Gibbs sampler \cite{michalak2003method} comes at the cost of generality by requiring the random fields of interest to be Gaussian. MCMC with the delayed rejection adaptive Metropolis (DRAM) sampling \cite{haario2006dram} is slightly more efficient and does not require the Gaussianity assumption; it has been used in experimental design for source identification~\cite{zhang2015efficient}, and is deployed as part of our algorithm. Gradient-based MCMC methods, such as hybrid Monte Carlo (HMC) sampling \cite{barajas2019efficient}, increase the slow convergence of these and other MCMC variants. However, the repeated computation of gradients of a Hamiltonian can be prohibitively expensive for high-dimensional transport problems.

With an exception of the method of distribution~\cite{boso-2020-data, boso-2020-learning}, the computational cost of Bayesian methods for data assimilation and statistical inference is dominated by multiple runs of a forward transport model. The computational burden can be significantly reduced by deploying a surrogate model, which provides a low-cost approximation of its expensive physics-based counterpart. Examples of such surrogates include polynomial chaos expansions \cite{zhang2015efficient, ciriello2019distribution} and Gaussian processes~\cite{elsheikh2014efficient, zhang2016adaptive}.  A possible surrogate-introduced bias can be reduced or eliminated altogether by the use of a two-stage MCMC~\cite{zhang2016adaptive}. Both polynomial chaos expansions and Gaussian processes suffer from the so-called curse of dimensionality, which refers to the degradation of their performance as the number of random inputs becomes large. 

Artificial neural networks in general, and deep neural networks in particular, constitute surrogates that remain robust for large numbers of inputs and outputs \cite{mo2018deep, mo2019deep}. Their implementations in open-source software offer an added benefit of being portable to advanced computer architectures, such as graphics processing units and tensor processing units, without significant input from the user. Our algorithm employs a convolutional neural network (CNN) as a surrogate, the role that is related to but distinct from other uses of neural networks in scientific computing, e.g., their use as a numerical method for solving differential equations \cite{lee1990neural, lagaris1998artificial}. 

In Section~\ref{sec:prbform} we formulate the problem of contaminant source identification from sparse and noisy measurements of solute concentrations. Section~\ref{sec:methods} contains a description of our algorithm, which combines MCMC with DRAM sampling (Section~\ref{sec:Bayes}) and a CNN-based surrogate of the forward transport model (Section~\ref{sec:cnn}). Results of our numerical experiments are reported in Section~\ref{sec:results}; they demonstrate that our method is about 20 times faster than MCMC with a physics-based transport model. Main conclusions drawn from this study are summarized in Section~\ref{sec:concl}.

\section{Problem Formulation}
\label{sec:prbform}

Vertically averaged hydraulic head distribution $h(\mathbf x)$ in an aquifer $\Omega$ with hydraulic conductivity $K(\mathbf x)$ and porosity $\theta(\mathbf x)$ is described by a two-dimensional steady-state groundwater flow equation,
\begin{linenomath*}
\begin{equation}
    \nabla \cdot (K \nabla h) = 0, \qquad \mathbf x \in \Omega,
    \label{eqa:flow}
    \end{equation}
\end{linenomath*}
subject to appropriate boundary conditions on the simulation domain boundary $\partial \Omega$. Once~\eqref{eqa:flow} is solved, average macroscopic flow velocity $\mathbf u(\mathbf x) = (u_1,u_2)^\top$ is evaluated as
\begin{linenomath*}
\begin{equation} \label{eqa:velo}
    \mathbf{u} = -\frac{K}{\theta} \nabla h.
    \end{equation}
\end{linenomath*}

Starting at some unknown time $t_0$ a contaminant with volumetric concentration $c_\text{s}$ enters the aquifer through point-wise or spatially distributed sources $\Omega_\text{s} \subset \Omega$. The contaminant continues to be released for unknown duration $T$ with unknown intensity $q_\text{s}(\mathbf x, t)$ (volumetric flow rate per unit source volume), such that $q_\text{s}(\mathbf x, t) \neq 0$ for $t_0 \le t \le t_0 + T$.
The contaminant, whose volumetric concentration is denoted by $c(\mathbf x,t)$, migrates through the aquifer and undergoes (bio)geochemical transformations with a rate law $R(c)$. Without loss of generality, we assume that the spatiotemporal evolution of $c(\mathbf x,t)$ is adequately described by an advection-dispersion-reaction equation,
\begin{linenomath*}
\begin{equation}\label{eqa:trans}
    \frac{\partial \theta c}{\partial t} = \nabla \cdot (\theta \mathbf{D}\nabla c) - \nabla \cdot(\theta \mathbf{u}c) - R(c) + q_\text{s} c_\text{s}, \qquad \mathbf x = (x_1,x_2)^\top \in \Omega, \quad t > t_0,
    \end{equation}
\end{linenomath*}
although other, e.g., non-Fickian, transport models \cite{neuman2009perspective, srinivasan2010random, severino2012lagrangian} can be considered instead. If the coordinate system is aligned with the mean flow direction, such that $\mathbf u = (u \equiv | \mathbf u |, 0)^\top$, then the dispersion coefficient tensor $\mathbf D$ in~\eqref{eqa:trans} has components
\begin{linenomath*}
\begin{equation}\label{eqa:disper}
      D_{11} = \theta D_\text{m} + \alpha_L u, \qquad 
      D_{22} = \theta D_\text{m} + \alpha_T u, \qquad
      D_{12} = D_{21} = \theta D_\text{m},
\end{equation}
\end{linenomath*}
where $D_\text{m}$ is the contaminant's molecular diffusion coefficient in water; and $\alpha_L$ and $\alpha_T$ are the longitudinal and transverse dispersivities, respectively.

Our goal is to estimate the location and strength of the source of contamination, $r(\mathbf x,t) = q_\text{s}(\mathbf x, t) c_\text{s}(\mathbf x, t)$,  by using the transport model~\eqref{eqa:flow}--\eqref{eqa:disper} and concentration measurements $\bar{c}_{m,i} =  \bar{c}(\mathbf{x}_m,t_i)$  collected at locations $\{\mathbf{x}_m\}_{m = 1}^{M}$ at times $\{t_i\}_{i = 1}^{I}$. The concentration data are corrupted by random measurement errors, such that
\begin{linenomath*}
\begin{equation}\label{eq:data}
\bar{c}_{m,i}  = c(\mathbf{x}_m,t_i) + \epsilon_{mi}, \qquad m = 1,\cdots,M, \quad i = 1,\cdots,I;
    \end{equation}
\end{linenomath*}
where $c(\mathbf{x}_m,t_i)$ are the model predictions, and the errors $\epsilon_{mi}$ are zero-mean Gaussian random variables with covariance $\mathbb{E}[\epsilon_{mi}\epsilon_{nj}] = \delta_{ij} R_{mn}$.
Here, $\mathbb{E}[\cdot]$ denotes the ensemble mean; $\delta_{ij}$ is the Kronecker delta function; and $R_{mn}$, with $m, n \in [1,M]$, are components of the $M\times M$ spatial covariance matrix $\mathbf{R}$ of measurements errors, taken to be the identity matrix multiplied by the standard deviation of the measurement errors. This model assumes both the model~\eqref{eqa:flow}--\eqref{eqa:disper} to be error-free and the measurements errors to be uncorrelated in time but not in space.

\section{Methods}
\label{sec:methods}

Our algorithm comprises MCMC with DRAM sampling and a CNN-based surrogate of the transport model~\eqref{eqa:flow}--\eqref{eqa:disper}. These two components are described below.

\subsection{MCMC with DRAM Sampling}
\label{sec:Bayes}

Upon a spatiotemporal discretization of the simulation domain, we arrange the uncertain (random) input parameters in~\eqref{eqa:flow}--\eqref{eqa:disper} into a vector $\mathbf m$ of length $N_m$; these inputs may include the spatiotemporally discretized source term $r(\mathbf x,t)$,  initial concentration $c_\text{in}(\mathbf x)$, hydraulic conductivity $K(\mathbf x)$, etc.
Likewise, we arrange the random measurements $\bar{c}_{m,i}$ into a vector $\mathbf d$ of length $N_d$, and the random measurement noise $\epsilon_{mi}$ into a vector $\boldsymbol{\varepsilon}$ of the same length. Then, the error model~\eqref{eq:data} takes the vector form
\begin{linenomath*}
\begin{equation}\label{eq:data-vec}
\mathbf d = \mathbf g(\mathbf m) + \boldsymbol{\varepsilon},
\end{equation}
\end{linenomath*}
where $\mathbf g(\cdot)$ is the vector, of length $N_d$, of the correspondingly arranged stochastic model predictions $c(\mathbf{x}_m,t_i)$ predicated on the model inputs $\mathbf m$.

In Bayesian inference, the parameters $\mathbf m$ are estimated probabilistically from both model predictions and (noisy) measurements by means of the Bayes theorem,
\begin{linenomath*}
\begin{equation}\label{eqa:likelihood}
f_{\mathbf m | \mathbf d} (\tilde{\mathbf m}; \tilde{\mathbf d}) = \frac{f_\mathbf{m}(\tilde{\mathbf m}) f_{\mathbf d | \mathbf m}(\tilde{\mathbf m}; \tilde{\mathbf d})}{f_\mathbf{d}(\tilde{\mathbf d})}, \qquad f_\mathbf{d}(\tilde{\mathbf d}) = \int f_\mathbf{m}(\tilde{\mathbf m}) f_{\mathbf d | \mathbf m}(\tilde{\mathbf m}; \tilde{\mathbf d}) \text d \tilde{\mathbf m}.
\end{equation}
\end{linenomath*}
Here, $\tilde{\mathbf d}$ is the deterministic coordinate in the phase space of the random variable $\mathbf d$; $f_\mathbf{m}$ is a prior probability density function (PDF) of the inputs $\mathbf m$, which encapsulates the information about the model parameters and contaminant source before any measurements are assimilated; $f_{\mathbf m | \mathbf d}$ is the posterior PDF of $\mathbf m$ that represents refined knowledge about $\mathbf m$ gained from the data $\mathbf d$; $f_{\mathbf d | \mathbf m}$ is the likelihood function, i.e., the joint PDF of concentration measurements conditioned on the corresponding model predictions that is treated as a function of $\mathbf m$ rather than $\mathbf d$; and $f_\mathbf{d}$, called ``evidence'', serves as a normalizing constant that ensures that $f_{\mathbf m | \mathbf d}(\mathbf m; \cdot)$ integrates to 1. Since $\boldsymbol{\varepsilon}$ in~\eqref{eq:data} or~\eqref{eq:data-vec} is multivariate Gaussian, the likelihood function has the form
\begin{linenomath*}
\begin{equation}
f_{\mathbf d | \mathbf m}(\tilde{\mathbf m}; \tilde{\mathbf d}) = \frac{1}{(2\pi)^{d/2}|\mathbf R|^{1/2}} \exp\left(-\frac{1}{2} \mathbf v^\top \mathbf R^{-1}\mathbf v\right), \qquad \mathbf v = \tilde{\mathbf d} - \mathbf g(\mathbf m).
\end{equation}
\end{linenomath*}

In high-dimensional nonlinear problems (i.e., problems with large $N_m$), such as~\eqref{eqa:flow}--\eqref{eqa:disper}, the posterior PDF $f_{\mathbf d | \mathbf m}$ cannot be obtained analytically and computation of the integral in the evidence $f_\mathbf{d}$ is prohibitively expensive. Instead, one can use MCMC to draw samples from $f_\mathbf{m}(\tilde{\mathbf m}) f_{\mathbf d | \mathbf m}(\tilde{\mathbf m}; \tilde{\mathbf d})$, without computing the normalizing constant $f_\mathbf{d}$.  A commonly used MCMC variant relies on the Metropolis--Hastings sampling \cite{gamerman2006markov}; this approach uses a zero-mean Gaussian PDF with tunable variance $\sigma^2$ to generate proposals near a previous sample, which are accepted with the acceptance rate given by the relative posterior value. The performance of the Metropolis-Hastings sampling depends on the choice of hyperparameters, such as $\sigma^2$, and  on how well the proposal PDF matches the target PDF. The choice of an inappropriate proposal PDF might cause an extremely slow convergence. 

We deploy the DRAM sampling---specifically its numerical implementation in \cite{miles2019pymcmcstat}---to accelerate the convergence of MCMC. It differs from the Metropolis--Hasting sampling in two aspects. First, the \emph{delayed rejection} \cite{green2001delayed} refers to the strategy in which a proposal's rejection in the first attempt is tied to the subsequent proposal that can be accepted with a combined probability for the two proposals; this rejection delay is iterated multiple times in the sampling process. Second, \emph{adaptive Metropolis} \cite{haario2001adaptive} uses past sample chains to tune the proposal distribution in order to accelerate the convergence of MCMC. The DRAM sampling is more efficient than other sampling strategies for many problems, including that of source identification~\cite{zhang2015efficient}.

\subsection{Deep Convolutional Neural Networks}
\label{sec:cnn}

Any MCMC implementation requires many solves of the transport model~\eqref{eqa:flow}--\eqref{eqa:disper} for different realizations of the input parameters $\mathbf m$. We use a CNN surrogate model to alleviate the cost of each solve. Several alternative input-output frameworks to construct a surrogate model are shown in Table~\ref{tab:in-out}. Among these, autoregressive models predict a concentration map only for the next time step. When measurements are collected at multiple times, an autoregressive model has to be repeatedly evaluated, for each realization of the inputs $ \mathbf{m}$. If the release time, conductivity field, and porosity are known, then $\mathbf{m}$ represents the initial concentration field $c_\text{in}(\mathbf{x})$. Otherwise, $\mathbf{m}$ is the stack of the maps of $c_\text{in}(\mathbf{x})$, conductivity field $K(\mathbf{x})$, porosity field $\theta(\mathbf{x})$, etc.

\begin{table}[htbp]
\caption{Alternative input-output frameworks for construction of surrogate models. The data are collected at $M$ locations $\mathbf x_m$ ($m=1,\cdots,M$) at $I$ times $t_i$ ($i=1,\cdots,I$). }
\label{tab:in-out}
\centering
\begin{tabular}{l l l c}
\hline
  Model & Input & Output & Modeling frequency \\
\hline
  PDE model                                 & $ \mathbf{m}$ & $\{c(\mathbf x,t_i)\}$ & 1  \\
  Image-to-image                          & $\mathbf{m}$ & $\{c(\mathbf x,t_i)\}$  & 1 \\
  Image-to-sensors                       & $\mathbf{m}$ & $\{c(\mathbf x_{m},t_i)\}$   & 1  \\
  Autoregressive image-to-image & $c(\mathbf{x},t)$ & $c(\mathbf{x},t + \Delta t)$  & $I$   \\
\hline
\end{tabular}
\end{table}

We choose an image-to-image regression model, rather than the autoregressive surrogate used in \cite{mo2019deep} to solve a similar source identification problem, for the following reasons. First, it is better at generalization than image-to-sensors models. Second, although autoregressive surrogates excel at regression tasks~\cite{mo2019deep}, they might become computationally expensive when the measurement frequency is high. 

\begin{figure}[htbp]
\centering
\includegraphics[width=0.9\textwidth]{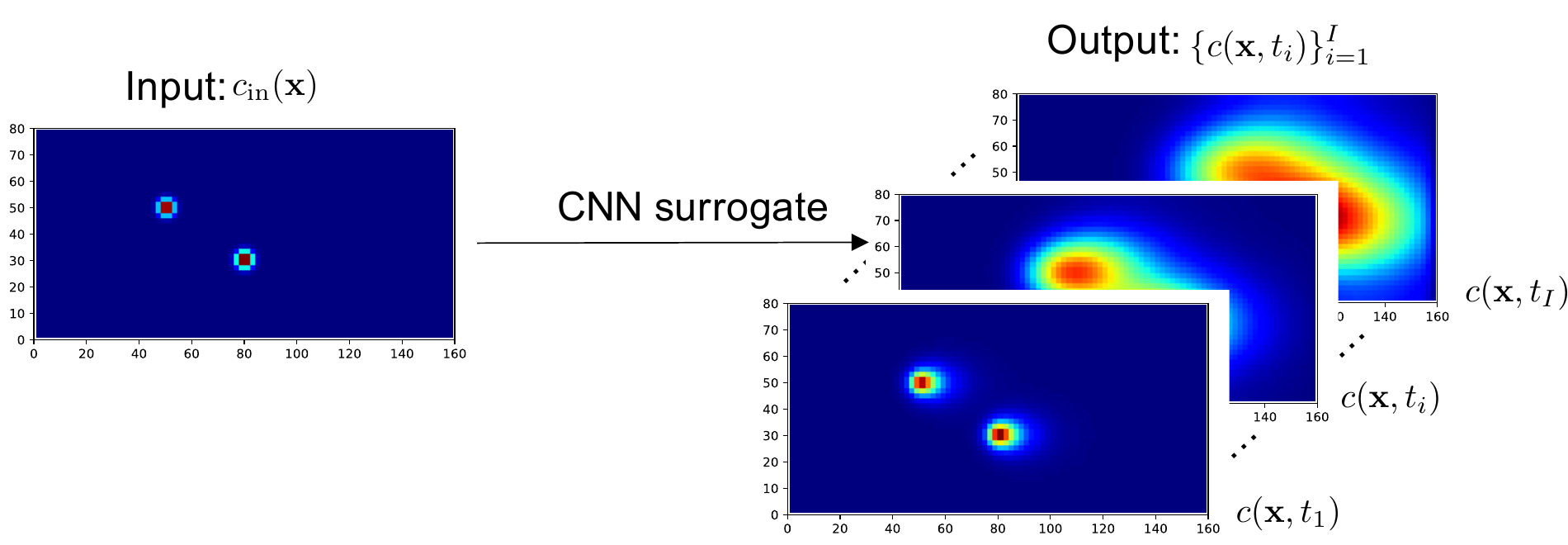}
\caption{A surrogate model constructed with a convolution neural network (CNN). The surrogate takes as input a set of uncertain parameters $\mathbf m$, e.g., an initial contaminant concentration field $c_\text{in}(\mathbf x)$ and returns as output temporal snapshots of the solute concentrations $c(\mathbf x, t_i)$ in an aquifer.}
\label{fig:in_out_2d}
\end{figure}

Our image-to-image regression model replaces the PDE-based transport model~\eqref{eqa:flow}--\eqref{eqa:disper} or $\mathbf g(\mathbf m)$ with a CNN $\mathbf N(\mathbf m)$ depicted in Figure~\ref{fig:in_out_2d}, i.e.,
\begin{linenomath*}
\begin{equation}\label{eqa:surro}
\mathbf g: \mathbf m \xrightarrow{\text{PDEs}} \{c(x_{m},t_i)\}_{m,i = 1}^{M,I} \;\text{is replaced with}\;
\mathbf{N}: \mathbf m \xrightarrow{\text{CNN}} \{c(\mathbf{x},t_i)\}_{i = 1}^{I}. 
\end{equation}
\end{linenomath*}
We start by attempting to demystify neural networks, which are spreading virally throughout the hydrologic community. A simplest way to relate the model output $\mathbf d$ to the model input $\mathbf m$ without having to run the model $\mathbf g$ is to replace the latter with a linear input-output relation $\hat{\mathbf{d}} = \mathbf W \mathbf m$, where $\mathbf W$ is an $N_d \times N_m$ matrix of weights whose numerical values are obtained by minimizing the discrepancy between the $\hat{\mathbf{d}}$ and $\mathbf d$ values that are either measured or computed with the model $\mathbf g$ or both. The performance of this linear regression, in which the bias parameters are omitted to simplify the presentation, is likely to be suboptimal, since a relationship between the inputs and outputs is likely to be highly nonlinear. Thus, one replaces $\hat{\mathbf{d}} = \mathbf W \mathbf m$ with a nonlinear model $\hat{\mathbf{d}} = \sigma(\mathbf W \mathbf m)$, in which a prescribed function $\sigma(\cdot)$ operates on each element of the vector $\mathbf W \mathbf m$. Examples of this so-called activation function include a sigmoidal function (e.g., $\tanh$) and a rectified linear unit (ReLU). The latter is defined as $\sigma(s) = \max(0, s)$, it is used here because of its current popularity in the field. The nonlinear regression model $\hat{\mathbf{d}} = \sigma(\mathbf W \mathbf m) \equiv (\sigma \circ \mathbf{W})(\mathbf m)$ constitutes a single ``layer'' in a network.

A (deep) fully connected neural network $\mathbf{N_f}$ comprising $N_l$ ``layers'' is constructed by a repeated application of the activation function to the input, 
\begin{subequations}\label{eqa:fc_net}
\begin{linenomath*}
\begin{equation}
\mathbf d = \mathbf{N_f}(\mathbf m; \boldsymbol\Theta) \equiv (\sigma_{N_l} \circ \mathbf{W}_{N_l - 1})\circ \ldots \circ (\sigma_{2} \circ \mathbf{W}_{1})(\mathbf m).
\end{equation}
\end{linenomath*}
In general, different activation functions might be used in one network. The parameter set $\boldsymbol\Theta = \{ \mathbf W_1,\ldots,\mathbf W_{N_l-1} \}$ consists of the weights $\mathbf W_n$ connecting the $n$th  and $(n+1)$st layers. In this recursive relation, 
\begin{linenomath*}
\begin{align}
\begin{cases}
    \mathbf{s}_1 = (\sigma_{2} \circ \mathbf{W}_{1})(\mathbf m)  \equiv \sigma_2(\mathbf{W}_{1} \mathbf m), \\
    \mathbf{s}_2 = (\sigma_{3} \circ \mathbf{W}_{2})(\mathbf{s}_1)  \equiv \sigma_3(\mathbf{W}_{2} \mathbf{s}_1), \\
    \vdots \\
    \mathbf d = (\sigma_{N_l} \circ \mathbf{W}_{N_l-1})(\mathbf{s}_{N_l - 2}) \equiv \sigma_{N_l}(\mathbf{W}_{N_l -1} \mathbf{s}_{N_l - 2}),
\end{cases}
\label{eqa:fc}
\end{align}
\end{linenomath*}
\end{subequations}
the weights $\mathbf W_1$ form a $d_1 \times N_m$ matrix, $\mathbf W_2$ is a $d_2 \times d_1$ matrix, $\mathbf W_3$ is a $d_3 \times d_2$ matrix,$\cdots$, and $\mathbf{W}_{N_l-1}$ is a $N_d \times d_{N_l-2}$ matrix. The integers $d_1,\cdots,d_{N_l-2}$ represent the number of neurons in the corresponding inner layers of the network. The fitting parameters $\boldsymbol\Theta$ are obtained, or the ``network is trained'', by minimizing the discrepancy between the prediction and the output in the dataset.

The size of the parameter set $\boldsymbol\Theta$ grows rapidly with the number of layers $N_l$ and the number of neurons $d_n$ in each inner layer. When the output layer contains hundreds or thousands of variables (aka ``features'', such as concentrations at observation wells collected at multiple times), this size can be unreasonably large. By utilizing a convolution-like operator to preserve the spatial correlations in the input, CNNs reduce the size of $\boldsymbol\Theta$ and scale much better with the number of parameters than their fully connected counterparts. CNNs are widely used to perform image-to-image regression. Details about a convolutional layer are not main concern of this study; we refer the interested reader to \cite{Goodfellow-et-al-2016} for an in-depth description of CNNs. In this study, a CNN is trained to predict the concentration map at times when the measurements were obtained.

Specifically, we use a convolutional encoder-decoder network to perform the regression with a coarse-refine process. In the latter, the encoder extracts the high-level coarse features of the input maps, and the decoder refines the coarse features to the full maps again \cite[fig.~2]{mo2019deep}. The $L_1$-norm loss function, $L_2$-norm weight regularization, and stochastic gradient descent \cite{bottou2010large} are used in the parameter estimation process.

It is worthwhile emphasizing that unlike some surrogate models, e.g., polynomial chaos which can predict a solution at any time, the CNN used in this study predicts only concentration maps for a short period. The reason is that for the inverse problem under consideration, only observations at measurement times are of interests and a model's ability to predict concentrations at later times is immaterial.

\section{Numerical Experiments}
\label{sec:results}

We use the CNN-based MCMC with DRAM sampling to identify a contamination source from sparse concentration measurements. A PDE-based transport model used to generate synthetic data is formulated in Section~\ref{sec:example}. Its CNN-based surrogate is developed and analyzed in Section~\ref{sec:cnn_example}. The performance of our approach in terms of the accuracy and efficiency vis-\`a-vis the PDE-based MCMC with DRAM sampling is discussed in Section~\ref{sec:performance}. 

\subsection{Contaminant Transport Model}
\label{sec:example}

\begin{figure}[htbp]
\centering
\includegraphics[width=0.9\textwidth]{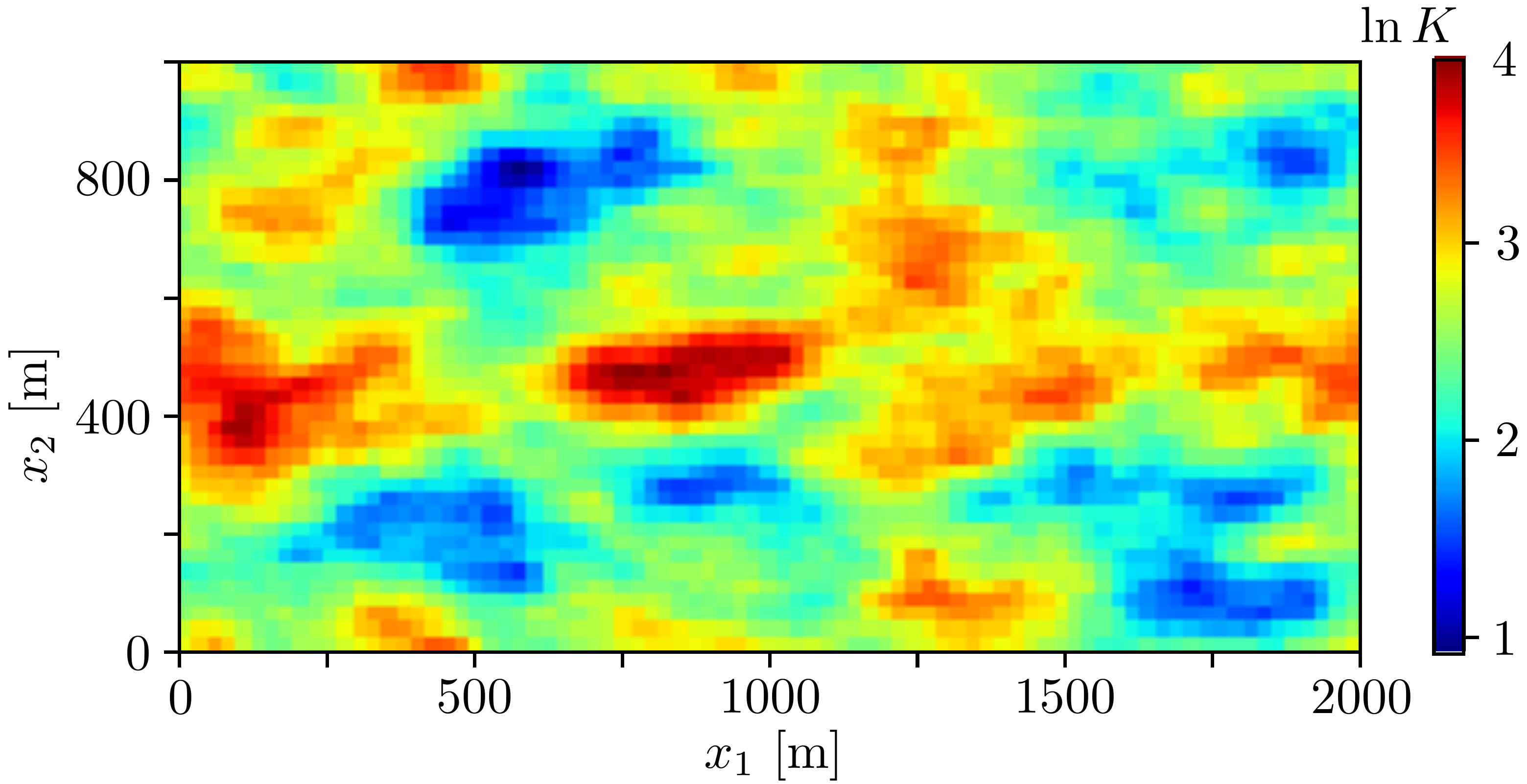}
\caption{Hydraulic conductivity $K(\mathbf x)$ [m/d], in logarithm scale.}
\label{fig:log_k}
\end{figure}

Our solute transport model consists of~\eqref{eqa:flow}--\eqref{eqa:disper} with $R(c) = 0$. A spatially varying hydraulic conductivity field $K(\mathbf x)$ is shown in Figure~\ref{fig:log_k} for a $1000$~m by $2000$~m rectangular simulation domain  discretized into $41 \times 81$ cells. We use the fast Fourier transform (see Algorithm 3 in \cite{lang2011fast}) to generate $K(\mathbf x)$ as a rescaled realization of the zero-mean multivariate Gaussian random field with the two-point covariance function 
\begin{linenomath*}
\begin{equation*} C(\mathbf x, \mathbf y) = \int_{\mathbb{R}^2} \text{e}^{-2\pi i \langle \mathbf p, \mathbf x - \mathbf y\rangle } | \mathbf p |^{-7/4} \text d p_1 \text d p_2, 
\end{equation*}
\end{linenomath*}
where $\langle \cdot,\cdot \rangle$ represents the Euclidean inner product on $\mathbb{R}^2$, and $\mathbf p=(p_1,p_2)^\top$.  

Porosity $\theta$ and dispersivities $\lambda_L$ and $\lambda_T$ are constant. The values of these and other flow and transport parameters, which are representative of a sandy alluvial aquifer in Southern California~\cite{liggett2015exploration,liggett2014fully}, are summarized in Table~\ref{tab:flow-trans}. Equation~\eqref{eqa:disper} is used to obtain the dispersion coefficients.

We consider an instantaneous, spatially distributed contaminant release taking place at time $t_0 = 0$. This replaces the source term $r(\mathbf x,t) = q_\text{s}(\mathbf x, t) c_\text{s}(\mathbf x, t)$ in~\eqref{eqa:trans} with the Dirac-delta source $r(\mathbf x,t) = r(\mathbf x) \delta (t)$  or, equivalently, with an unknown initial contaminant distribution $c_\text{in}(\mathbf x)$. Our goal is to reconstruct the latter from the noisy concentration data $\bar c_{m,i}$  collected at $M = 20$ locations $\{\mathbf{x}_m\}_{m = 1}^{M}$ at  $\{t_i\}_{i = 1}^{I} = \{3, 4,\ldots,18)$ years after the contaminant release ($I = 16$).  

\begin{figure}[htbp]
\centering
\includegraphics[width=0.9\textwidth]{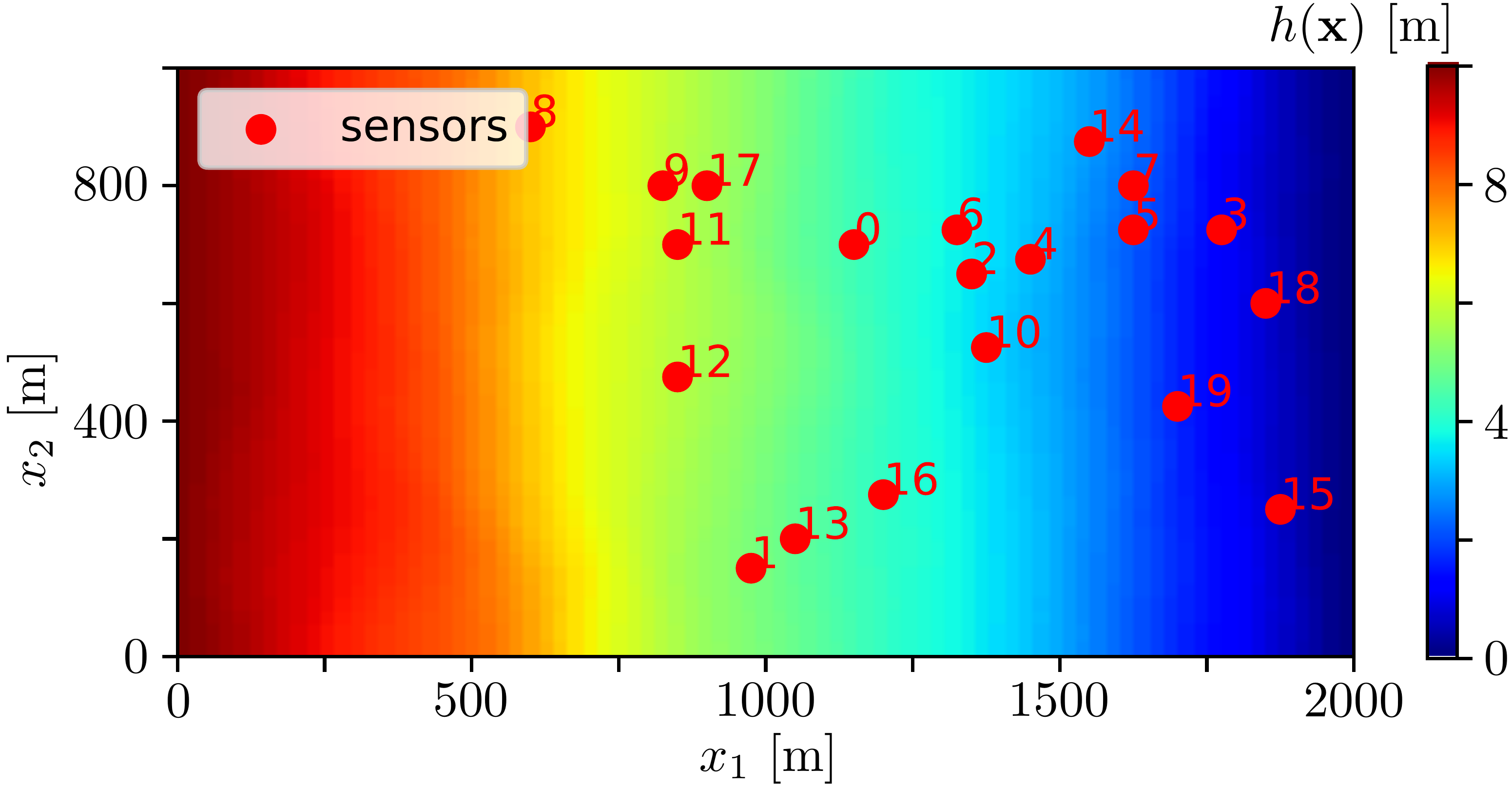}
\caption{Hydraulic head distribution $h(\mathbf x)$ [m] and locations of 20 observational wells. The flow is driven by constant heads $h_L = 10$~m and $h_R = 0$ maintained at the left and right boundaries, respectively; no-flow boundary conditions are assigned to the upper and lower boundaries.}
\label{fig:head}
\end{figure}

\begin{table}[htbp]
\caption{Values of hydraulic and transport parameters, which are representative of sandy alluvial aquifers in Southern California~\cite{liggett2015exploration,liggett2014fully}. }
\label{tab:flow-trans}
\centering
\begin{tabular}{l c c}
\hline
  Parameter & Value & Units \\
\hline
  Porosity, $\theta$ & $0.3$  &  $-$\\
  Molecular diffusion, $D_\text{m}$ &  $10^{-9}$ & m$^2$/d  \\
  Longitudinal dispersivity, $\alpha_L$ & $10$ & m\\
  Dispersivity ratio, $\alpha_L / \alpha_T$ & $10$  & $-$ \\
\hline
\end{tabular}
\end{table}

We used \texttt{Flopy} \cite{bakker2016scripting}, a \texttt{Python} implementation of MODFLOW \cite{harbaugh2005modflow} and MT3DMS \cite{bedekar2016mt3d}, to solve the flow~\eqref{eqa:flow} and transport~\eqref{eqa:trans} equations, respectively. With constant hydraulic head values on the left and right boundaries, the head distribution $h(\mathbf x)$ is shown in Figure~\ref{fig:head}, together with the locations of 20 observational wells.

\begin{figure}[htbp]
\includegraphics[width=\textwidth]{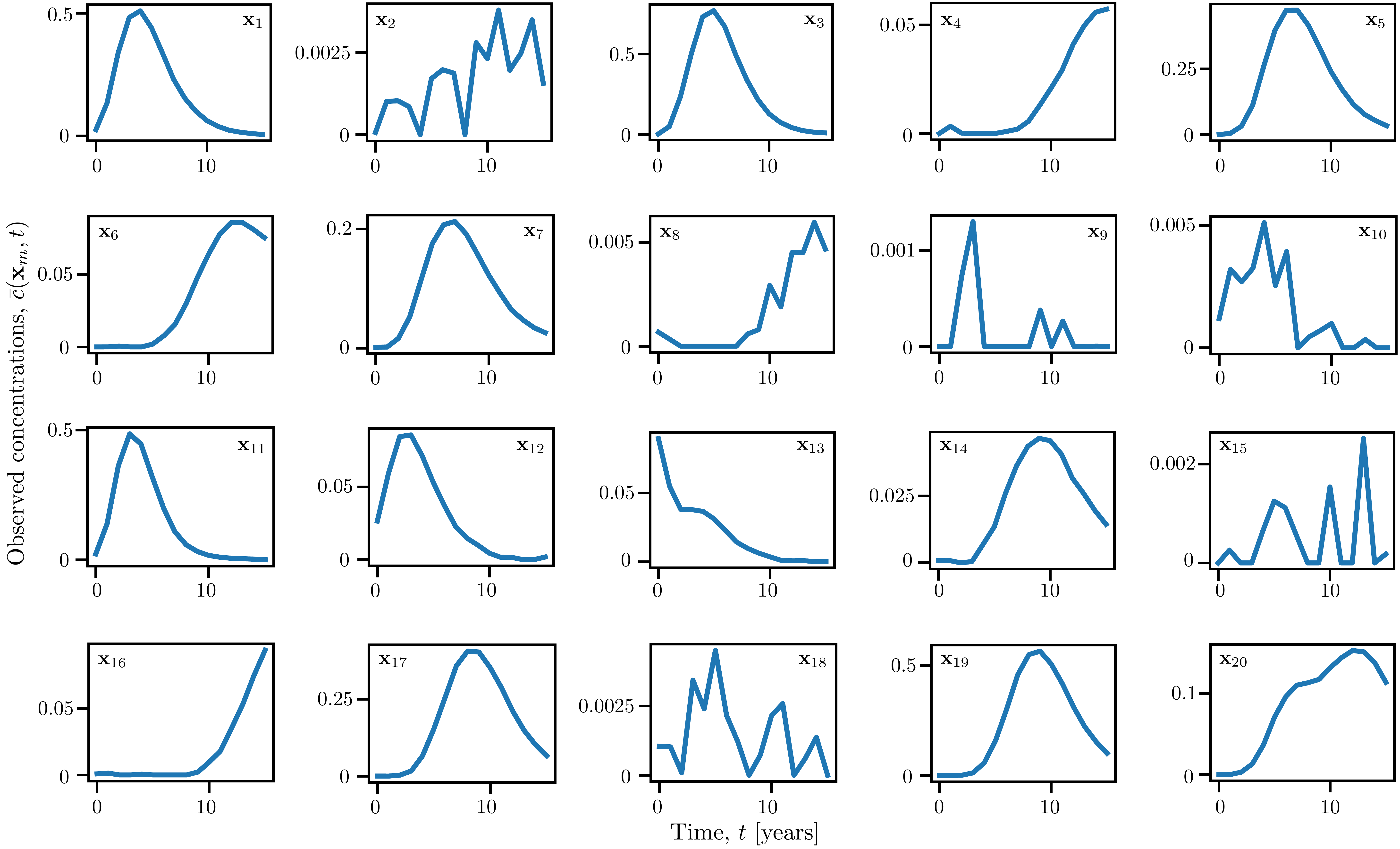}
\caption{Contaminant breakthrough curves $c(\mathbf x_m,t)$ observed in the wells whose locations $\mathbf x_m$ ($m=1,\ldots,20$) are shown in Figure~\ref{fig:head}.}
\label{fig:bt_curves}
\end{figure}

The initial contaminant distribution consists of $N_p$ co-mingling Gaussian plumes,
\begin{linenomath*}
\begin{equation}\label{eqa:gaussian_plumes}
c_\text{in}(x_1,x_2) =  \sum^{N_p}_{i = 1} S_i \exp\left[- \frac{(x_1 - x_{1,i})^2 + (x_2 - x_{2,i})^2}{2\sigma_i^2} \right],
\end{equation}
\end{linenomath*}
each of which has the strength $S_i$ and the width $\sigma_i$, and is centered at the point $(x_{1,i}, x_{2,i})$. The true, yet unknown, values of these parameters are collated in Table~\ref{tab:prior_ref} for $N_p = 2$; they are used to generate the measurements $\bar c_{m,i}$ by adding the zero-mean Gaussian noise with standard deviation $\sigma_{\epsilon} = 0.001$. These data form the $20$ breakthrough curves shown in Figure~\ref{fig:bt_curves}. 

\begin{table}
\caption{Prior uniform distributions for the meta-parameters $\mathbf m$ characterizing the initial contaminant plume~\eqref{eqa:gaussian_plumes},  and the true, yet unknown, values of these parameters.}
\label{tab:prior_ref}
\centering
\begin{tabular}{l c c c c c c c c}
\hline
   & $x_{1,1}$ & $x_{2,1}$ & $x_{1,2}$ & $x_{2,2}$ & $S_1$ & $\sigma_1$ & $S_2$ & $\sigma_2$\\
\hline
  Interval  & [0,700] & [50,900] & [0,700] & [50,900] & [0,100] & [13,20] & [0,100] & [13,20] \\
  Truth & $325$ & $325$ & $562.5$ & $625$ & $30$ & $15$ & $50$ & $17$   \\
\hline
\end{tabular}
\end{table}

The lack of knowledge about the initial contaminant distribution $c_\text{in}(\mathbf x)$ is modeled by treating these parameters, $\mathbf{m} = (x_{1,i}, x_{2,i}, \sigma_{i}, S_i)$ with $i = 1$ and $2$, as random variables distributed uniformly on the intervals specified in Table~\ref{tab:prior_ref}. These uninformative priors are refined as the measurements are assimilated into the model predictions.

\subsection{Construction and Accuracy of CNN Surrogate}
\label{sec:cnn_example}

As discussed in Section~\ref{sec:methods}, although only model predictions at 20 wells are strictly necessary for the inversion, the use of full concentration distributions $c(\mathbf x, t_i)$ as output of the CNN-based surrogate has better  generalization properties. We used $N = 1600$ solutions (Monte Carlo realizations) of the PDE-based transport model~\eqref{eqa:trans} for different realizations of the initial condition $c_\text{in}(\mathbf x)$ to train the CNN; another $N_\text{test} = 400$ realizations were retained for testing. These 2000 realizations of the initial concentration $c_\text{in}(\mathbf x)$ in~\eqref{eqa:gaussian_plumes} were generated with Latin hyper-cube sampling of the uniformly distributed input parameters $\mathbf m$ from Table~\ref{tab:prior_ref}. The CNN contains three dense blocks with $N_l = 6, 12, 6$ internal layers, uses a growth rate $R_g = 40$, and has $N_\text{in} = 64$  initial features; it was trained for 300 epochs. The CNN's  output is 16 stacked maps of the solute concentration $c(\mathbf x, t_i)$ at $t_i = (3, 4,\ldots, 18)$ years after the contaminant release.

\begin{figure}[htbp]
\centering
\includegraphics[width=\textwidth]{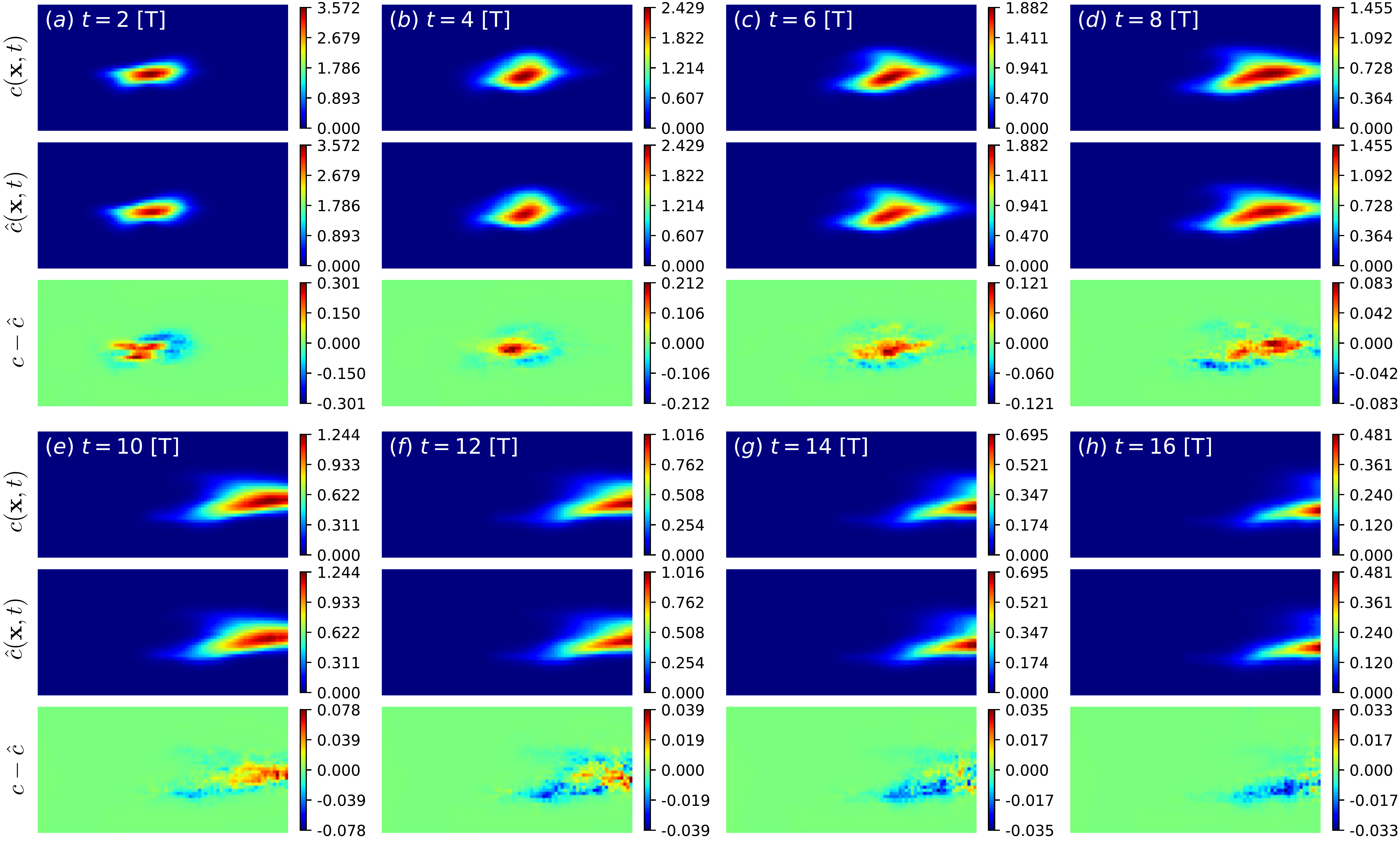}
\caption{Temporal snapshots of the solute concentration alternatively predicted with the transport model ($c$, top row) and the CNN surrogate ($\hat c$, second row) for a given realization of the initial concentration $c_\text{in}(\mathbf x)$. The bottom row exhibits the corresponding errors of the CNN surrogate, $(c-\hat c)$. The times in the upper left corner correspond to the number of years after contaminant release. }
\label{fig:test_nn}
\end{figure}

Figure~\ref{fig:test_nn} exhibits temporal snapshots of the solute concentrations alternatively predicted with the transport model, $c(\mathbf x, t_i)$, and the CNN surrogate, $\hat c(\mathbf x,t_i)$, for a given realization of the initial concentration $c_\text{in}(\mathbf x)$ at eight different times $t_i$. The root mean square error of the CNN surrogate, $\| c(\mathbf x, t_i) - \hat c(\mathbf x, t_i) \|_2$, falls to $0.023$ at the end of the training process. It is worthwhile emphasizing here that the $N = 1600$ Monte Carlo realizations used to train our CNN surrogate are but a small fraction of the number of forward solves needed by MCMC.

\subsection{MCMC Reconstruction of Contaminant Source}
\label{sec:performance}

\begin{figure}[htbp]
\centering
\includegraphics[width=\textwidth]{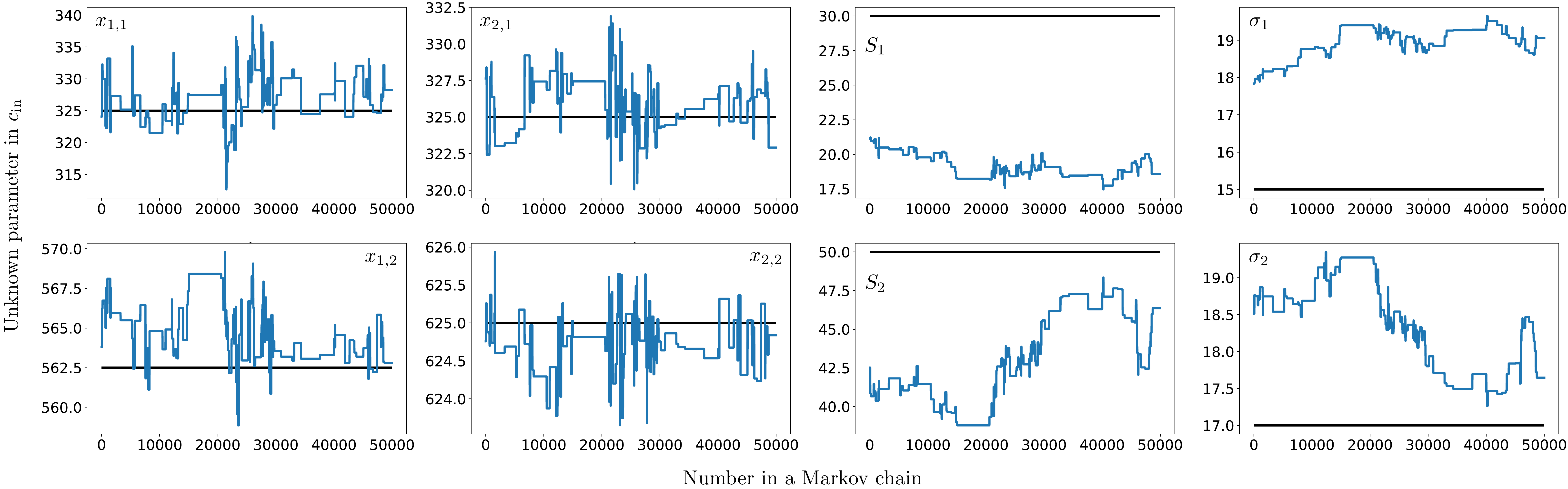}
\caption{MCMC chains of the parameters $\mathbf m$ characterizing the initial plume configuration $c_\text{in}(\mathbf x)$ obtained by sampling from the transport model~\eqref{eqa:trans}. Each Markov chain represents a parameter value plotted as function of the number of iterations (links in the chain). The black horizontal lines are the true values of each parameter.}
\label{fig:PDE_trace}
\end{figure}

We start by analyzing the performance of MCMC with DRAM sampler of $\mathbf m$ when the PDE-based transport model~\eqref{eqa:trans} is used to generate realizations of $c(\mathbf x,t_i)$. Since the model is treated as exact, this step allows us to establish the best plume reconstruction provided by our implementation of MCMC. The latter relied on $100000$ samples of $\mathbf m$, the first half of which was used in the ``burn-in'' stage and, hence, are not included into the estimation sample set. Figure~\ref{fig:PDE_trace} exhibits sample chains for each of the six parameters $\mathbf m$ characterizing the initial plume configuration $c_\text{in}(\mathbf x)$. Visual inspection of these plots reveals that MCMC does a good job identifying the centers of mass of the two co-mingling plumes, $(x_{1,i}, x_{2,i})$ with $i= 1$ and $2$; identification of the spatial extent, $\sigma_i$, and strength, $S_i$, of these plumes is less accurate.

\begin{table}[htbp]
\caption{MCMC chain statistics---mean, standard deviation, integrated autocorrelation time $\tau$, and Geweke convergence diagnostic $p$---of the parameters $\mathbf m$ characterizing the initial plume configuration $c_\text{in}(\mathbf x)$ obtained by sampling from the PDE model. Also shown is the total contaminant mass of the two co-mingling plumes, $M_1$ and $M_2$.}
\label{tab:chain_PDE}
\centering
\begin{tabular}{l c c c c c}
\hline
   Parameter  & True value & Mean & Std & $\tau$ & $p$\\
\hline
  $x_{1,1}$   & $325$   & $327.5836$ & $3.3924$ & $1046.3394$ & $0.9991$  \\
  $x_{2,1}$   & $325$   & $325.7773$ & $1.6108$ & $1289.5577$ & $0.9929$  \\
  $x_{1,2}$   & $562.5$ & $564.3320$ & $1.9967$ & $2218.9018$ & $0.9881$  \\
  $x_{2,2}$   & $625$   & $624.7743$ & $0.3203$ & $402.0658$  & $0.9998$  \\
  $S_1$       & $30$    & $18.6853$  & $0.5007$ & $1713.8339$ & $0.9699$  \\
  $\sigma_1$  & $15$    & $19.1371$  & $0.2365$ & $2172.9087$ & $0.9837$  \\
  $S_2$       & $50$    & $44.3071$  & $2.8493$ & $4441.9589$ & $0.7632$  \\
  $\sigma_2$  & $17$    & $18.0939$  & $0.5932$ & $4409.0626$ & $0.8832$  \\
  $M_1$       & $20.4244$ & $20.6709$&    $-$     & $-$       & $-$\\
  $M_1$       & $43.5802$& $43.74$ &    $-$     & $-$       & $-$\\
\hline
\end{tabular}
\end{table}

Table~\ref{tab:chain_PDE} provides a more quantitative assessment of the performance of the PDE-based MCMC. The standard deviations of the MCMC estimates of the plumes' centers of mass, $(x_{1,i}, x_{2,i})$, is no more than $1\%$ of their respective means, indicating high confidence in the estimation of these key parameters. The standard deviations of the other parameter estimates, relative to their respective means, are appreciably higher. Also shown in Table~\ref{tab:chain_PDE} are Sokal's adaptive truncated periodogram estimator of the integrated autocorrelation time $\tau$ \cite{sokal1997monte}, and the Geweke convergence diagnostic $p$ \cite{geweke1991evaluating}. These quantities are routinely used to diagnose the convergence of Markov chains. The former provides an average number of dependent samples in a chain that contain the same information as one independent sample; the latter quantifies the similarity between the first $10\%$ samples and the last $50\%$ samples.

Although somewhat less accurate, the estimates of the spatial extent, $\sigma_i$, and strength, $S_i$, of the co-mingling plumes is more than adequate for field applications.  Their estimation errors cannot be eliminated with more computations, as suggested by a very large number of samples used in our MCMC. Instead, they reflect the relative dearth of information provided by a few sampling locations.

\begin{figure}[htbp]
\centering
\includegraphics[width=\textwidth]{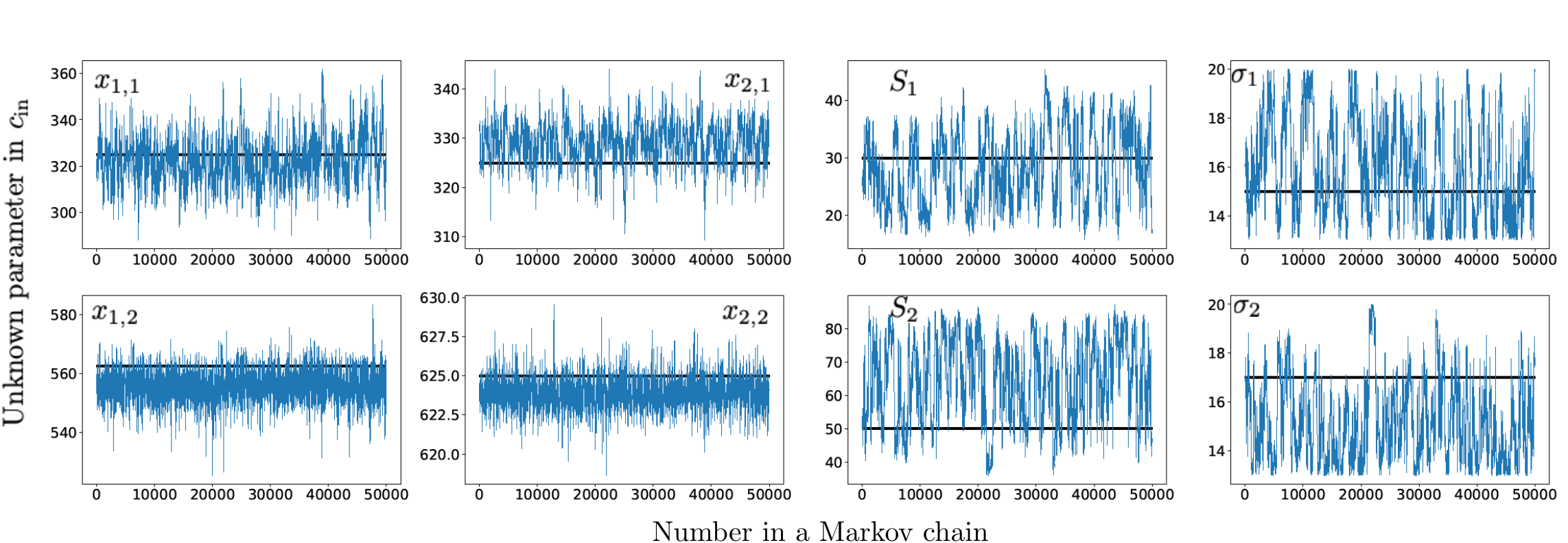}
\caption{MCMC chains of the parameters $\mathbf m$ characterizing the initial plume configuration $c_\text{in}(\mathbf x)$ obtained by sampling from the CNN surrogate~\eqref{eqa:fc_net}. Each Markov chain represents a parameter value plotted as function of the number of iterations (links in the chain). The black horizontal lines are the true values of each parameter.}
\label{fig:nn_trace_short}
\end{figure}

\begin{table}[htbp]
\caption{MCMC chain statistics---mean, standard deviation, integrated autocorrelation time $\tau$, and Geweke convergence diagnostic $p$---of the parameters $\mathbf m$ characterizing the initial plume configuration $c_\text{in}(\mathbf x)$ obtained by sampling from the CNN surrogate. Also shown is the total contaminant mass of the two co-mingling plumes, $M_1$ and $M_2$. }
\label{tab:chain_NN}
\centering
\begin{tabular}{l c c c c c}
\hline
   Parameter & True value & Mean & Std & $\tau$ & $p$\\
\hline
  $x_{1,1}$   & $325$   & $322.3274$ & $9.8827$ & $189.8946$ & $0.9944$  \\
  $x_{2,1}$   & $325$   & $328.8859$ & $3.9956$  & $231.9033$ & $0.9992$  \\
  $x_{1,2}$   & $562.5$ & $555.4074$ & $4.3167$ & $35.8577$  & $0.9983$  \\
  $x_{2,2}$   & $625$   & $623.8933$ & $0.8944$  & $43.2115$  & $0.9999$  \\
  $S_1$       & $30$    & $28.4441$  & $6.4531$ & $514.4594$ & $0.8100$  \\
  $\sigma_1$  & $15$    & $15.9822$  & $1.9291$  &$537.7868$ & $0.9094$  \\
  $S_2$       & $50$    & $64.6830$  & $12.1613$ &$540.6132$ & $0.9962$  \\
  $\sigma_2$  & $17$    & $15.1550$  & $1.6076$  &$543.3779$ &$0.9964$  \\
  $M_1$       & $20.4244$ & $21.9306$&    $-$     & $-$       & $-$\\
  $M_1$       & $43.5802$& $44.8789$&    $-$     & $-$       & $-$\\
\hline
\end{tabular}
\end{table}

\begin{figure}[htbp]
\centering
\includegraphics[width=\textwidth]{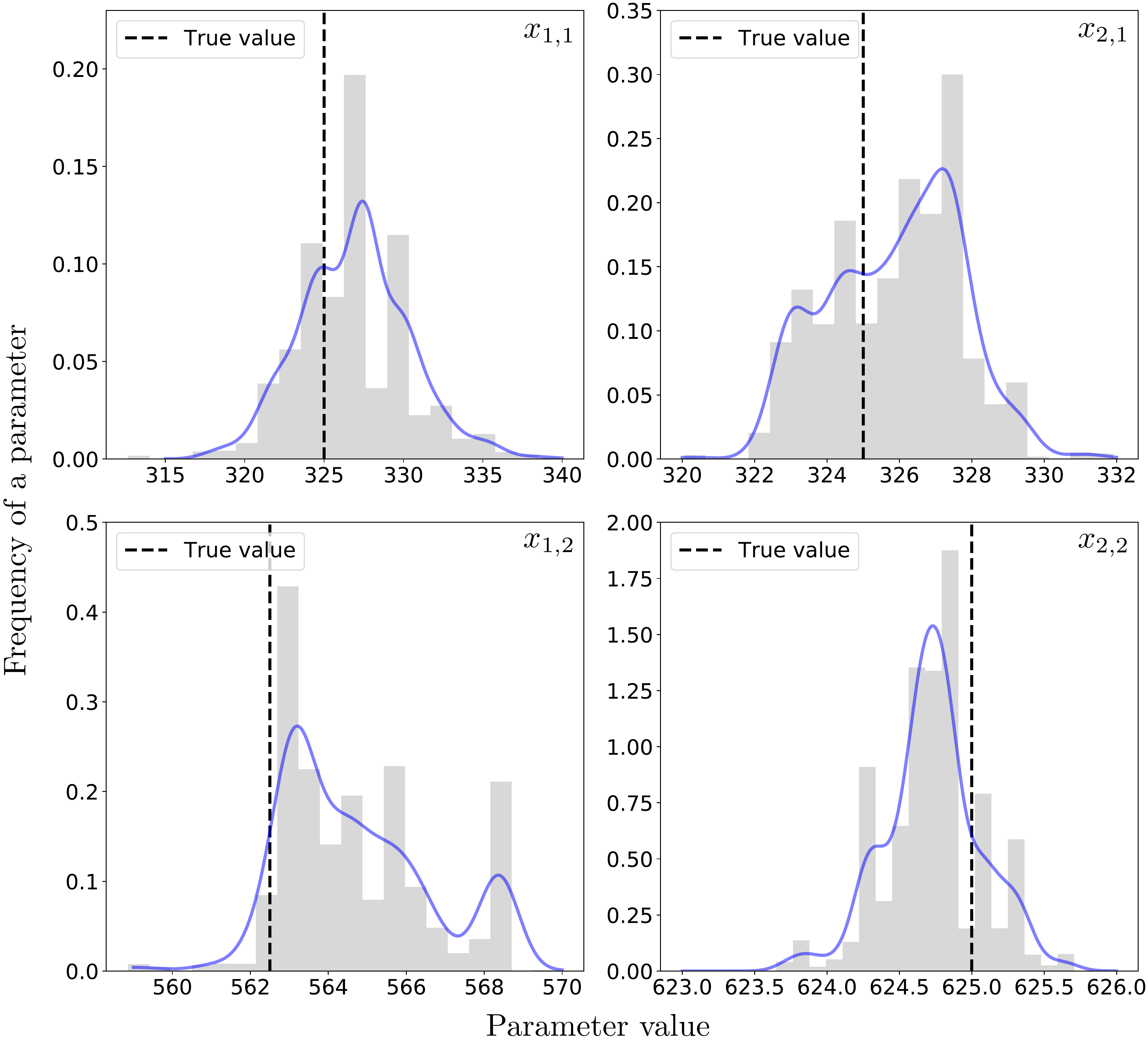}
\caption{Probability density functions (solid lines) and histograms (gray bars) of the centers of mass of the two commingling spills, $(x_{1,1}, x_{2,1})$ and $(x_{1,2}, x_{2,2})$, computed with MCMC drawing samples from the PDE-based transport model. Vertical dashed lines represent the true locations.}
\label{fig:pde_pdf}
\end{figure}

\begin{figure}[htbp]
\centering
\includegraphics[width=\textwidth]{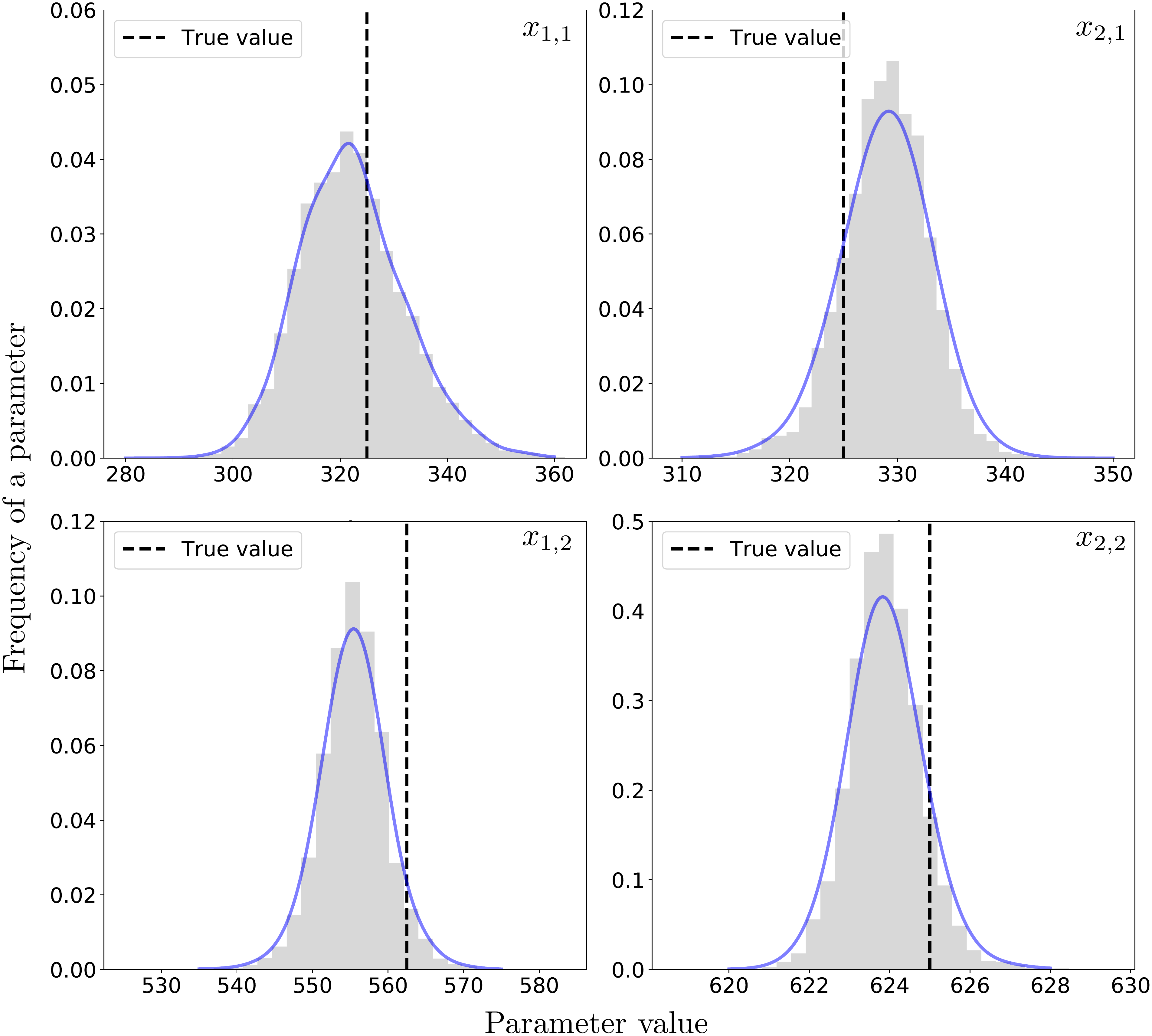}
\caption{Probability density functions (solid lines) and histograms (gray bars) of the centers of mass of the two commingling spills, $(x_{1,1}, x_{2,1})$ and $(x_{1,2}, x_{2,2})$, computed with MCMC drawing samples from the CNN surrogate. Vertical dashed lines represent the true locations.}
\label{fig:nn_pdf}
\end{figure}

Next, we repeat the MCMC procedure but using the CNN surrogate to generate samples. Figure~\ref{fig:nn_trace_short} exhibits the resulting MCMC chains of the parameters $\mathbf m$, i.e., the parameter values plotted as function of the number of samples $N$ (excluding the first $50000$ samples used in the burn-in stage). Because of the prediction error of the CNN surrogate, the chains differ significantly from their PDE-based counterparts in Fig.~\ref{fig:PDE_trace}. They are visibly better mixed, an observation that is further confirmed by the fact that the integrated autocorrelation times $\tau$ in Table~\ref{tab:chain_NN} are much smaller than those reported in Table~\ref{tab:chain_PDE}. 
However, the standard deviations (Std) for the parameter estimators are slightly larger than those obtained with the PDE-based MCMC; this implies that the CNN prediction error undermines the ability of  MCMC to narrow down the posterior distributions. The posterior PDFs for the centers of mass of the two commingling plumes, $(x_{1,i}, x_{2,i})$, are shown in Figs.~\ref{fig:pde_pdf} and~\ref{fig:nn_pdf}. The discrepancy between the actual and reconstructed (as the means of these PDFs) locations is within 7~m; it is of negligible practical significance.

Comparison of Tables~\ref{tab:chain_PDE} and~\ref{tab:chain_NN} reveals that, similar to the PDE-based sampler, the CNN-based sampler provides more accurate estimates of the source location  $(x_{1,i}, x_{2,i})$ than of its spread ($\sigma_i$) and strength ($S_i$). However, in practice, one is more interested in the total mass of the released contaminant ($M$) rather than its spatial configuration (characterized by $\sigma_i$ and $S_i$). The mass of each of the commingling plumes in~\eqref{eqa:gaussian_plumes}, $M_1$ and $M_2$, is  
\begin{linenomath*}
\begin{equation}\label{eqa:total_mass}
M_i = \theta \int_{\Omega_i} c_\text{in}(\mathbf{x}) \text d \mathbf x, \qquad \Omega_i:[x_{1,i} \pm 100] \times [x_{2,i} \pm 100], \qquad i=1,2.
\end{equation}
\end{linenomath*}
Both the PDE- and CNN-based MCMC strategies yield accurate estimates of $M_1$ and $M_2$ (Tables~\ref{tab:chain_PDE} and~\ref{tab:chain_NN}).

\subsection{Computational Efficiency of MCMC with CNN Surrogate}

Our CNN-based MCMC is about 20 times faster than MCMC with the PDE-based transport model (Table~\ref{table:cost}). This computational speed-up can be attributed to either the algorithmic improvement or the different hardware architecture or both. That is because while the off-the-shelf PDE-based software utilizes central processing units (CPU), NN training takes place on graphics processing units (GPUs), e.g., within the \texttt{GoogleColab} environment used in our study, without much effort on the user's part. To disentangle these sources of computational efficiency, we also run the CNN-based MCMC on the same CPU architecture used for the PDE-based MCMC. Table~\ref{table:cost} demonstrates that the CNN-based MCM ran on CPU is about twice faster than the PDE sampler. This indicates that the computational speed-up of the CNN-based sampler is in large part due to the use of GPUs for CNN-related computations. One could rewrite PDE-based transport models to run on GPUs, but it is not practical. At the same time, no modifications or special expertise are needed to run the \texttt{Pytorch}  implementation~\cite{paszke2019pytorch} of neural networks on GPUs.

\begin{table}[htbp]
\caption{Total run time (in seconds) of the MCMC samplers, $T_\text{run}$, based on the PDE-based transport model and its CNN surrogate. The PDE sampler uses CPUs; the CNN sampler is trained and simulated on GPUs provided by \texttt{GoogleColab}; for the sake of comparison, also reported is the run time of the CNN sampler on the CPU architecture used for the PDE-based sampler. In all three cases,  MCMC consists of $N_\text{sam} = 10^5$ samples. The average run-time per sample, $T_\text{ave}$, is defined as $T_\text{ave} = (T_\text{run} + T_\text{train}) / N_\text{sum}$, where $T_\text{train}$ is the CNN training time.\label{table:cost}}
\label{tab:comp_cost}
\centering
\begin{tabular}{l c c c}
\hline
   & $T_\text{run}$ & $T_\text{train}$ & $T_\text{ave}$\\
\hline
  PDE                 & $101849.0$ & $0$          & $1.01849$  \\
  CNN on GPU   & $1101.7$     & $4007.4$ &  $0.05109$   \\
  CNN on CPU   & $ 37450.0 $ & $4007.4$ & $0.41457$ \\
\hline
\end{tabular}
\end{table}

\section{Conclusions and Discussion}
\label{sec:concl}

We proposed an MCMC approach that uses DRAM sampling and draws samples from a CNN surrogate of a PDE-based model. The approach was used to reconstruct contaminant release history from sparse and noisy measurements of solute concentration. In our numerical experiments, water flow and solute transport take place in a heterogeneous two-dimensional aquifer; the goal is to identify the spatial extent and total mass of two commingling plumes at the moment of their release into the aquifer.  Our analysis leads to the following major conclusions.
\begin{enumerate}
    \item The CNN-based MCMC is able to identify the locations of contaminant release, as quantified by the centers of mass of commingling spills forming the initial contaminant plume.
    \item Although somewhat less accurate, the estimates of the spread and strength of these  spills is adequate for field applications. Their integral characteristics, the total mass of each spill, are correctly identified.
    \item The estimation errors cannot be eliminated with more computations. Instead, they reflect both the ill-posedness of the problem of source identification and the relative dearth of information provided by sparse concentration data.
    \item Replacement of a PDE-based transport model with its CNN-based surrogate increases uncertainty in, i.e., widens the confidence intervals of, the source identification. 
    \item The CNN-based MCMC is about 20 times faster than MCMC with the high-fidelity transport model. This computational speed-up is in large part due to the use of GPUs for CNN-related computations, while the PDE solver utilizes CPUs.
\end{enumerate}

While we relied on a CNN to construct a surrogate of the PDE-based model of solute transport, other flavors of NNs could have be used for this purpose.  We are not aware of published comparisons of alternative NNs in the context of image-to-image prediction, which is required by our MLMC method. In the somewhat related context of spectrum sensing~\cite{ye2019comparison}, the comparison of a fully connected neural network (FNN), a recurrent neural network (RNN), and a CNN revealed the FNN to have small utility for ordered and correlated samples like images; the CNN and RNN to exhibit a comparable performance in terms of accuracy, and the  RNN to be more efficient in terms of memory requirements.

In general, the direct comparison of the performance of a FNN and a CNN on the same task is not helpful and can be misleading because of the freedom of the architecture of each network and the presence of multiple tuning parameters in both. However, the results reported in section~\ref{sec:cnn} suggest that a FNN would contain significantly more parameters given the size of the input and output images. This applies even to a relatively shallow FNN. Some studies in image classification, e.g.,~\cite{chen2017nb}, claim that, relative to FNNs, CNNs require more training data to achieve convergence and avoid overfitting. Even if this conclusion generalizes to our application it is of little practical significance, because we found the combined cost of the training-data generation and NN training to be significantly lower than the cost of MCMC sampling. 

Properly trained autoregressive models and RNNs can be a strong competitor to CNNs, because they perform like a fixed time-step predictor and, consequently, \emph{might} generalize better. RNNs are likely to be more expensive because of higher prediction frequency, but require less memory for each prediction. Our implementation of CNNs utilized a parallel GPU architecture to carry out convolutional operations. However, since GPUs become more affordable, this drawback can be ignored.

Our computational examples deal with an instantaneous contaminant release. Since a CNN has been shown to provide an accurate surrogate of the PDE-based transport model with temporally distributed sources~\cite{mo2019deep} and since MCMC is known to accurately reconstruct prolonged contaminant release history~\cite{barajas2019efficient}, our CNN-based MCMC is expected to provide comparable computational gains when used to identify prolonged contaminant releases.  

\begin{acknowledgements}
This work was supported in part by Air Force Office of Scientific Research under award  FA9550-18-1-0474, National Science Foundation under award 1606192, and by a gift from TOTAL. There are no data sharing issues since all of the numerical information is provided in the figures produced by solving the equations in the paper. We used the code from \cite{mo2019deep} to construct and train the convolutional neural network.
\end{acknowledgements}

%
%

\bibliographystyle{spmpsci}      

\bibliography{mcmc+cnn.bib}   


\end{document}